\pdfoutput=1

\documentclass{raa}
\usepackage{graphicx,times}
\usepackage{natbib}
\usepackage{amssymb,amsmath}

\usepackage{subfigure}
\usepackage{url}
\usepackage{CJK}

\renewcommand{\vec}[1]{ {\mathbf #1} }

\newcommand{\grad}{ {\bf \nabla } }

\newcommand{\Fig}{{Figure}}
\newcommand{\Figs}{{Figures}}

\bibpunct{(}{)}{;}{a}{}{,}

\begin{document}

\title{A Magnetic Bald-Patch Flare in Solar Active Region 11117}

\author{Chaowei Jiang \inst{1,2}
  \and Xueshang Feng \inst{2,1}
  \and S.~T. Wu \inst{3}
  \and Qiang Hu \inst{3}}
%% Here is an example of three authors come from different institutes.
%% For single author or all the authors from an institute, use "\inst{}" only

\institute{HIT Institute of Space Science and Applied Technology,
  Shenzhen, 518055, China; {\it
    chaowei@hit.edu.cn}\\
  \and SIGMA Weather Group, State Key Laboratory for Space Weather,
  Center for Space Science and Applied Research,
  Chinese Academy of Sciences, Beijing 100190, China\\
  \and Center for Space Plasma and Aeronomic Research, The University
  of Alabama in Huntsville, Huntsville, AL 35899, USA
  %% Please give the E-mail address of the author, to whom future
  %% correspondence and
  %% offprint requests will be sent.
  \vs \no {\small Received XXX; accepted XXX} }

\abstract{With SDO observations and a data-constrained MHD model, we
  identify a confined multi-ribbon flare occurred on 2010 October 25
  in solar active region 11117 as a magnetic bald patch (BP) flare
  with strong evidences. From the photospheric magnetic field observed
  by SDO/HMI, we find there is indeed magnetic BPs on the PILs which
  match parts of the flare ribbons. From the 3D coronal magnetic field
  derived from a MHD relaxation model constrained by the vector
  magnetograms, we find strikingly good agreement of the BP separatrix
  surface (BPSS) footpoints with the flare ribbons, and the BPSS
  itself with the hot flaring loop system. Moreover, the triggering of
  the BP flare can be attributed to a small flux emergence under the
  lobe of the BPSS, and the relevant change of the coronal magnetic
  field through the flare is well reproduced by the pre-flare and
  post-flare MHD solutions, which match the corresponding pre and
  post-flare AIA observations, respectively. Our work contributes to the study of
  non-typical flares that constitute the majority of solar flares but
  cannot be explained by the standard flare model.
  \keywords{Sun: flares --- Sun: corona --- Magnetic fields ---
    Magnetohydrodynamics (MHD) --- Methods: numerical} }

\authorrunning{C.-W. Jiang et al. }            %author_head in even pages
\titlerunning{A Bald-Patch Flare in Active Region 11117}  % title_head in odd pages
\maketitle

%________________________________________________ sections below
%

\section{Introduction}
\label{sec:intro}

%Basic theory of the flare and important relation with magnetic
%topology structure. The BP and its relation with flare.
It is commonly believed that solar flares are explosive manifestation
of magnetic energy release in the solar atmosphere by magnetic
reconnection \citep{Shibata2011}. The magnetic energy can be stored in
the solar corona by means of magnetic flux emergence and/or energy
injection by photospheric surface motions (shear/twist) from the sun's
interior, and fast magnetic reconnection is the core process that
release the stored energy. Depending on the specific coronal
conditions, flares may be related with or without coronal mass
ejections (CMEs). The eruptive flares, i.e., those accompanied with
CMEs, are typically observed as with two flare ribbons roughly
parallel along the main polarity inversion line (PIL) of magnetic
field on the photosphere. These classical two-ribbon flares have been
extensively studied,
%leading to a %explained well by a
converging to a well-known standard flare model \citep[i.e., the CSHKP
flare model,][]{Carmichael1964, Sturrock1966, Hirayama1974, Kopp1976},
in which an eruptive magnetic flux rope (MFR) rises above the PIL,
stretches the overlying field lines, produces a vertical current sheet
underneath and triggers reconnection there, which results in the
two-ribbon brightenings at the footprints of the reconnecting field
lines. Most of these flares occur in a common magnetic structure called
sigmoidal active regions, which is manifestation of sheared bipolar field and
favourable for formation of MFR.

On the other hand, there are numerous flares, either eruptive
or confined, occur in a much more complicated manner than the
classical two-ribbon flares. Flares associated with multiple ribbons
of complex pattern are frequently reported in recent observations,
such as quasi-circular ribbon flares \citep{Masson2009, WangH2012},
three-ribbon flares \citep{WangH2014}, X-shaped flares \citep{LiY2016, LiuR2016}
and remote
flare ribbons distinct from the eruptive core site
\citep[or the secondary ribbon, e.g.,][]{ZhangJ2014}. These atypical flares appear to be more difficult
to explain as no standard model exists. However, considering that a
very big portion of the flares are atypical, it is necessary to
emphasize on study of these complex flares \citep{Dalmasse2014}.

The variety of flare-ribbon pattern roots in the complexity of the
underlying, invisible coronal field. For instance, the closed
circular-like ribbon is found to be produced by a coronal magnetic
null-point configuration \citep{Masson2009, WangH2012,
  Jiang2013MHD}. To generally understand how the flares occur, it is
required to have the knowledge of where the reconnection might be
triggered in a given configuration of the coronal field. Under typical
coronal conditions, the plasma resistivity is extremely low and the
magnetic field is frozen into the plasma almost everywhere in solar
corona. As such, magnetic reconnection can only occur in certain
places where the current form thin layers for the resistivity to be
important to induce sufficient dissipation \citep{Demoulin2006,
  Demoulin2007}. Such thin layers include the magnetic separatrices
and more commonly, quasi-separatrix layers, across which the
connectivity of field lines discontinues or changes abruptly, and
consequently narrow, enhanced current sheets can easily form along
there due to the photospheric driving motions \citep{Priest2002,
  Titov2002, Longcope2005}.

%BPs are usually associated with magnetic reconnection because of the separatrices.
% Separatrices are an exception where thin current layers can be
%formed and then dissipated, rebuilding the magnetic field-line linkage
%(i.e., magnetic reconnection).

Theoretical studies show that there are interesting places where
coronal magnetic field lines become tangent to the photosphere at the
PIL \citep[e.g.,][]{Wolfson1989,Low1992,Titov1993}. In such places,
opposite to the normal case, the field line crosses the PIL from {\it
  negative} to {\it positive} polarity. These places are dubbed `bald
patches' (BPs) with the visual reference to a haircut with magnetic
field lines being associated to hairs \citep{Titov1993}. Since the
photosphere can be regarded as a line-tying boundary for the coronal
field, a BP field line is thus very special because it is anchored at
the BP in addition to its footpoints. As a result, the continuous
set of BP field lines
%that graze the photosphere at the BP
defines a separatrix surface (BPSS) of magnetic topology, across which
the field-line linkage is discontinuous. This is rather important
considering that BP is the only place that can define magnetic
separatrices beside the case in which magnetic nulls are present in
the corona. The field lines immediately above a BP are concave up
(i.e., forming magnetic dips) and can held material against solar
gravity, which thus usually implies the existence of filament
associated with BP \citep{Titov1999, Mackay2010}. The criterion for
the existence of BPs in the models of potential and linear force-free
fields was given in detail by \citet{Titov1993}.

% in which ways are the currents is formed and magnetic reconnection
% can take place,
For the case of BPSS, current sheets are very prone to be formed and
thus reconnection can be triggered by shearing motions at the
photosphere footpoints of the BP-separatrix field lines or pushing by
flux emergence under the BP-separatrix lobe. By a 3D numerical MHD
simulation driven by a bottom shearing velocity, \citet{Pariat2009}
shows that current can form all along the curved BP separatrices where
the reconnection can take place successively. There is another
possibility in which vertical current sheet can be formed just above
BP due to a converging and upward movement of the field lines from
different lobes \citep[see Figure~1 in][]{Titov1993}. By this pinching
effect, the two BP lobes are brought into contact and lead to a
vertical current sheet formed between the oppositely directed magnetic
fields, which eventually reconnect. Besides, during the evolution of
some magnetic configurations, BPs may be precursors of the emergence
of a null point in the coronal field \citep{Bungey1996}, being again
associated with reconnection
processes. % say sth about relation with flux rope
The BP separatrix also have a close relationship with MFR, which holds
a central position in many flare/CME models
\citep[e.g.,][]{Forbes2006, Torok2005, Kliem2006}. During a MFR
forming in the corona or emerging from below the photosphere, there
might be a state when the MFR is bodily attached at the photosphere
besides its two legs, and the BP separatrix is the interface between
the MFR and the ambient field \citep{Titov1999}. When viewed from
above, this MFR-BPSS usually exhibits an S or inverse-S shape, and
reconnection there can produce a hot plasma in corresponding field
lines, which can explain the observations of coronal sigmoids. Indeed,
\citet{Jiang2014NLFFF} recently found an almost coincidence of an EUV
sigmoid in AR~11283 with the BPSS of a corresponding MFR reconstructed
from HMI vector magnetogram.

%what's other's work on the evidence of BP and flare relations
In observation, BP was first related with flares by
\citet{Aulanier1998}, who found a close correspondence between the BP
separatrices and the H$\alpha$ and X-ray emissions in a very small
flare (or sub-flare) in AR~7722, which was firstly interpreted as a
so-called `BP flare'. Several observation studies also show that the
reconnection triggered at BP can be correlated to eruptive flares.
%due to BP separatrices, in which strong current sheets can be formed by
%photospheric motions or flux emergence and trigger reconnection.
\citet{Delannee1999} studied another example of BP-related
flare and CME near AR~8100 and AR~8102, and suggested that the
eruption is trigger by reconnection occurs in a vertical current
sheet, which is formed above the BP due to the photospheric shearing
motions and the line-tie condition at the BP
\citep[e.g.,][]{Titov1993}. \citet{WangTJ2002} showed that in AR~8210
the emerging motion of a twist flux rope may drives a slow
reconnection at a BP (manifesting as a flare), which removes the
overlaying flux confining the flux rope to give way for the flux rope
to expand and form a CME.
Besides, BP are associated with a wider
range of phenomena in the chromsphere and transition region, e.g., the
transition region brightenings \citep{Fletcher2001}, in surge
ejections and arch filament systems \citep{Mandrini2002}, and in
Ellerman bombs, i.e., small-scale transient H$\alpha$ brightenings
\citep{Pariat2004}. All these phenomena are closely related with flux
emergence, which is frequently observed in the solar atmosphere.

%
% Although the concept of BP-flare is proposed for many years, a
% concrete evidence is still lack because the following reasons: The
% shortage of more direct evidence of BP-related flare may be because
% the high-resolution vector magnetogram is lack and the related flare
% is rather small-scale.

However, it should be noted that, probably due to the lack of the
high-resolution vector magnetogram or the rather small scale of the
related flare, the evidence for existence of BP in aforementioned
examples of BP-flare was computed using an extrapolation of the
photospheric field, particularly, the linear extrapolations (e.g., the
potential, linear force-free or linear magnetohydrostatic models)
based only on the line-of-sight component of the photospheric field.
Such kind of evidence, obviously is very indirect and may be not
reliable because of the limitation of the models, considering that the
location of BPs can be deduced directly from vector magnetograms (with
the 180$^{\circ}$ ambiguity resolved). Also it is very difficult to
use a simple linear field model to recover the complex non-potential
coronal structures, which is generally very nonlinear
\citep[e.g.,][]{Wiegelmann2008}, although in some occasions, the basic
topology of the magnetic field can be `sketched' roughly using the
linear models \citep{Demoulin1997}. To reconstruct the coronal field
and capture the magnetic topology more precisely, it is necessary to
use more realistic models, like the nonlinear force-free field (NLFFF) extrapolation
or even the MHD model, with the photospheric vector field as input.

Now with the high-resolution and high-cadense vector magnetic field
data and EUV observations from SDO, we have more opportunity to test
the theory of BP-induced flare. Very recently, basing on a sophisticated
topological analysis of NLFFF extrapolation from HMI vector magnetogram,
\citet{LiuR2014} suggested that
magnetic reconnection at the BPSS may play an important role in
producing an unorthodox X-class long-duration confined flare in AR~11339.
%with direct evidence.
In this paper we revisit a C-class confined flare event in AR~11117, which has been used for our validation of
a CESE--MHD model for reconstructing coronal magnetic field based on the SDO/HMI vector magnetograms
\citep{Jiang2012c}. By analyzing the reconstructed magnetic
field, we have preliminarily shown that there is a relation between
this flare and a BP found on the vector magnetogram
near the flare site and time. Here with the same modeling data we will
give more concrete evidence of the correlation between the BP and the
flare, and further show how the magnetic reconnection at the BP is
activated and the corresponding change of the BP separatrices. The remainder of
the paper is organized as follows. In Section~\ref{sec:obser} we
describe the observations of the flare by SDO/AIA and in
Section~\ref{sec:topol} we study the magnetic field on the photosphere
and of the 3D topology, to provide the evidence of the BP-flare and
show how the flare is triggered. We finally discuss the results and
conclude in Section~\ref{sec:conc}.

\begin{figure}[htbp]
  \centering
  \includegraphics[width=0.8\textwidth]{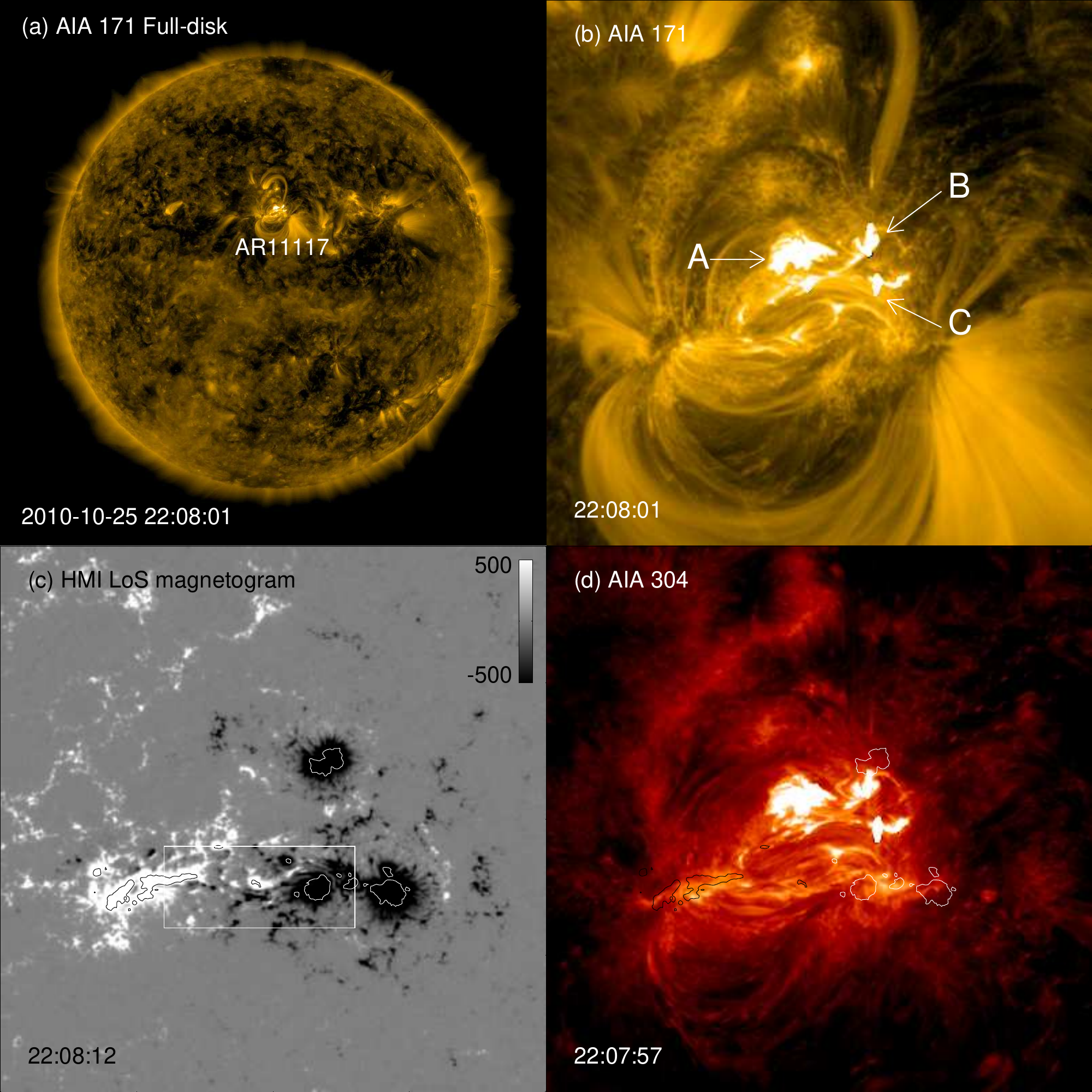}
  \caption{Location of the active region 11117 and the flare site. (a)
    Full-disk {\it SDO}/AIA 171 {\AA} image with AR 11117 outlined by
    the white rectangle on the image. (b) Enlarged view of AR11117 in
    the full-disk image. The
    flare site with mainly three brightening kernels are labeled as A, B
    and C. (c) The HMI line-of-sight magnetogram for the same
    field-of-view of (b). Contour lines are plotted at $\pm
    1000$~G. The box region denotes a fast flux emergence site. (d)
    the AIA 304 {\AA} image of the same region with the contour lines
    in (c) overlapped on the image. }
  \label{fig:flaresite}
\end{figure}

\begin{figure*}[htbp]
  \centering
  \includegraphics[width=\textwidth]{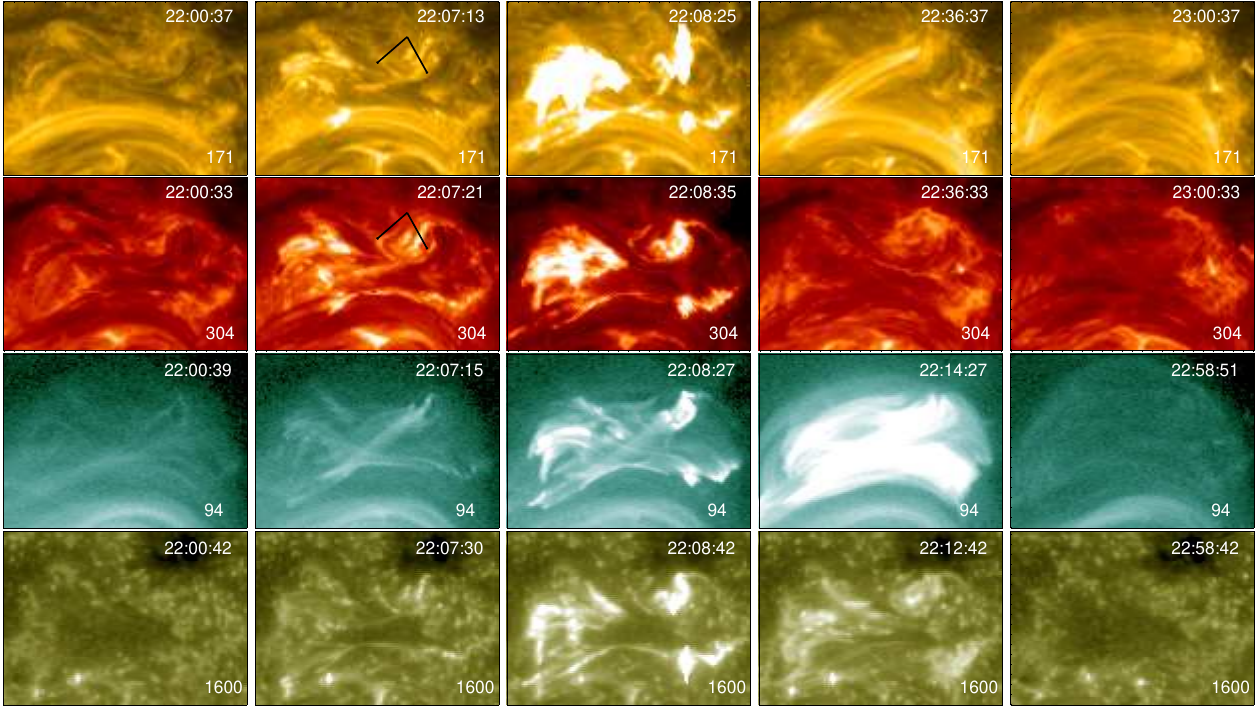}
  \caption{Coronal evolution through the flare observed by AIA in channels of 171~{\AA}, 304~{\AA}, 94~{\AA} and 1600~{\AA}.}
  \label{fig:flarelocal}
\end{figure*}

\section{Observations}
\label{sec:obser}

%Active region NOAA AR 11117 was observed by SDO from 2010 October 20
%to 2010 November 2, mainly during Carrington Rotation 2102. On 2010
%October 25, it was

As shown in \Fig~\ref{fig:flaresite}, the studied flare occurred in
AR~11117 on 2010 October 25, when the AR was crossing the central
meridian of the solar disk with latitude of $22^{\circ}$N. On this day,
AR~11117 is the largest one of the ARs on disk, and only one flare
with class of C2.3 was recorded according to the GOES (Geostationary
Operational Environmental Satellite) 1--8 {\AA} light curve, which
starts at 22:06~UT, peaks at 22:12~UT and ends at about 22:18 UT. The
AR consists of two leading sunspots of negative polarity, one in the
north and the other in the south, followed by an elongated positve
polarity in the east, and fast magnetic flux emergence to the core of
the region during the day (denoted by the box in \Fig~\ref{fig:flaresite}c). However, rather unexpectedly, the flare
did not happen in the flux emerging site with strong magnetic shear,
but in the intermediate region between the sunspots (especially much
closer to the north sunspot, which is further away from the flux
emergence site than the south one, see \Fig~\ref{fig:flaresite}b and d), where the field is relatively weak
and no distinct magnetic shear is observed.

% On this date solar activity was dominated by AR~11117 with many
% small B-class flares observed and near the end of the day, the C2.3
% flare        the central part of the active region shows distinct
% brightenings at the flare peak time.

% careful description of the AIA brightenings: the shape, the
% distributions is needed.

As recorded well by the AIA, the flaring process is confined in a
rather low altitude without inducing major changes in the coronal
loops or eruptions. \Fig~\ref{fig:flarelocal} gives the evolution of
the flare site in detail with four EUV channels, the 171~{\AA} for the
corona and hot loops shown by 94~{\AA}, the 304~{\AA} for the
upper chromosphere, and 1600~{\AA} for the lower/middle chromosphere. It begins to
brighten at 22:07~UT and ends at 22:15~UT. The flare brightening can
be seen most clearly for several minutes around the time of 22:08~UT,
which the \Fig~\ref{fig:flaresite} is plotted for. We only roughly
regards it consisting of three brightening kernels labeled as A, B and
C on the AIA image (see \Fig~\ref{fig:flaresite}(b)), since these
small patches are very close to each other and seem to be
connected. The brightening patch at the east, A, is much bigger than
the other two, B and C, at the west, indicating that more energy is
delivered at A, and by carefully inspecting time-series of
the AIA images (\Figs~\ref{fig:flarelocal}) or
time-lapse movies, it seems that the flare is first brightened at kernel A,
where it might be activated or triggered, and then brightened at B and
C, possibly because the energetic particles following the magnetic
field lines to their footpoints. Note that actually the flare patch A
consists of many even smaller patches by inspecting the 1600~{\AA},
indicating multiple complex footpoints of the coronal magnetic field
involved with reconnection.

As shown by the arrows on the AIA images of 171~{\AA} and 304~{\AA} in
\Fig~\ref{fig:flarelocal}, there appears to be a small curved
dark feature overlying the flare kernels, indicating that field lines
there exhibit some dips to support this filament-like material.  This
mini-filament sustained for hours before the onset of the flare, and
disappeared after the flare (see the AIA images at time of
23:00~UT). As can be seen in AIA 171~{\AA}, there is a new set of
coronal loops expanding after the flare, just in the same location of
the disappeared mini-filament.

\begin{figure*}[htbp]
  \centering
  \includegraphics[width=0.8\textwidth]{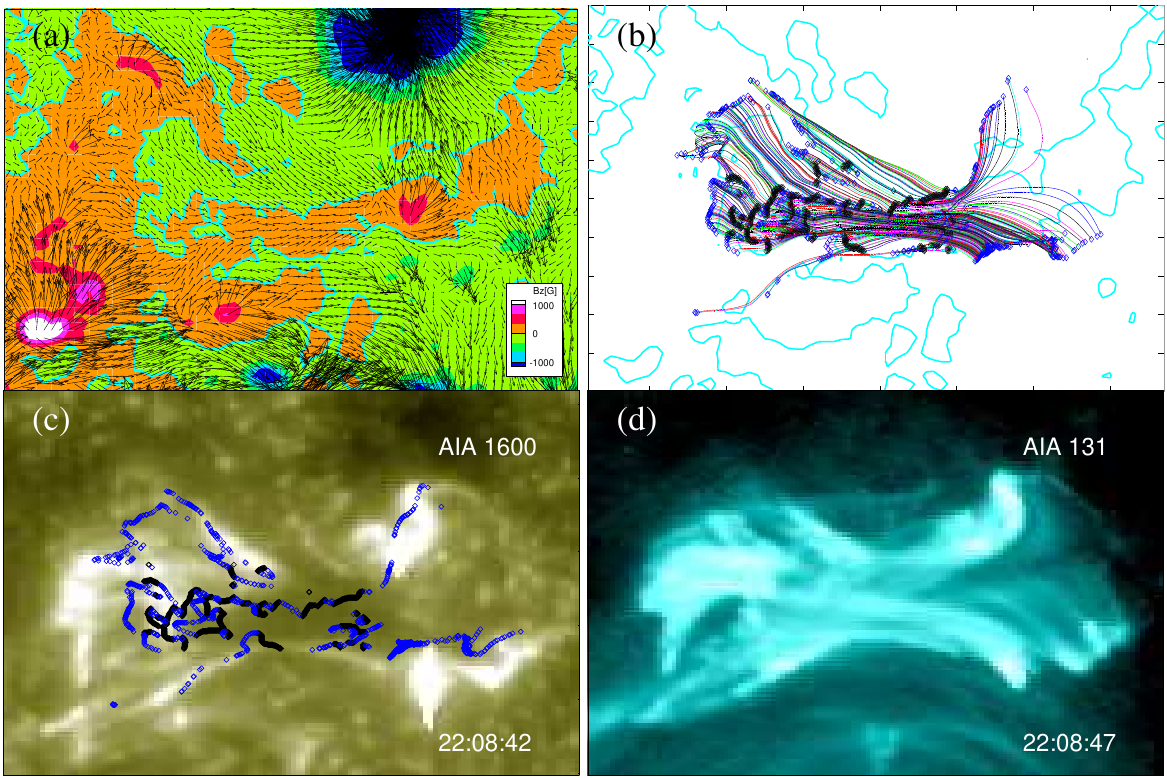}
  \caption{Spatial relation between the flare site and the BPSS.  (a)
    Magnetic vector fields at the photosphere taken by HMI at
    22:12~UT; (b) Magnetic field lines form the BPSS. The small
    diamond shapes in black denote the BPs, and those in blue for the
    footpoints of the BPSS field lines. The thick cyan line represents
    the photospheric magnetic PILs. (c) Flare brightening patches
    observed by AIA in 1600~{\AA}, and the overlaid are the BPs and
    BPSS footpoints. (d) Hot flaring loops observed by AIA in
    131~{\AA}.
    %A nice agreement of the BPs and BPSS footpoints with the flare patches and the BPSS with the flaring loops can be seen.
    }
  \label{fig:AIA_BP}
\end{figure*}

\begin{figure}[htbp]
  \centering
  \includegraphics[width=0.8\textwidth]{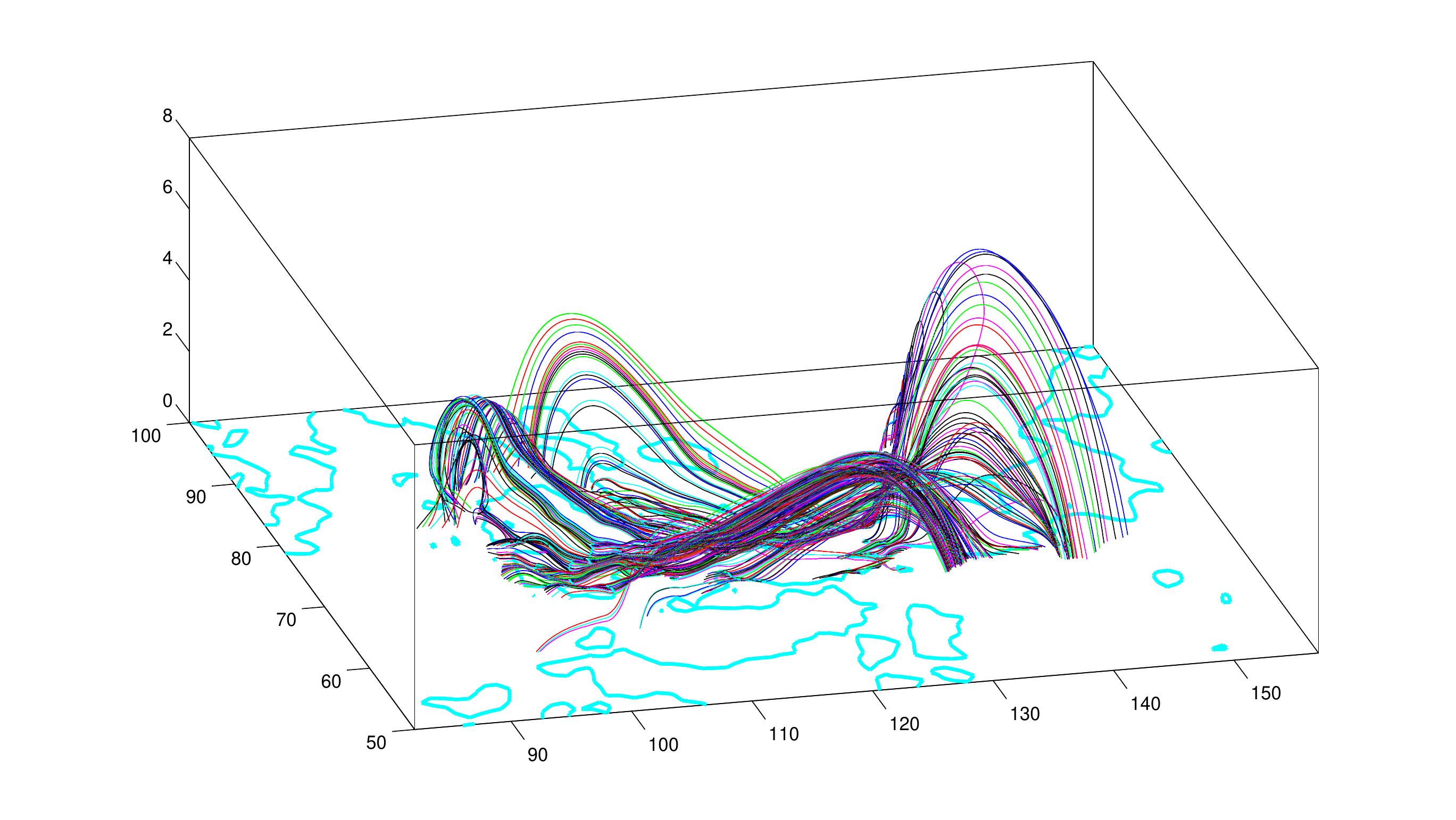}
  \caption{3D view of the BPSS field lines as shown in \Fig~\ref{fig:AIA_BP}(b).}
  \label{fig:BP3D}
\end{figure}

\begin{figure*}[htbp]
  \centering
  \includegraphics[width=0.8\textwidth]{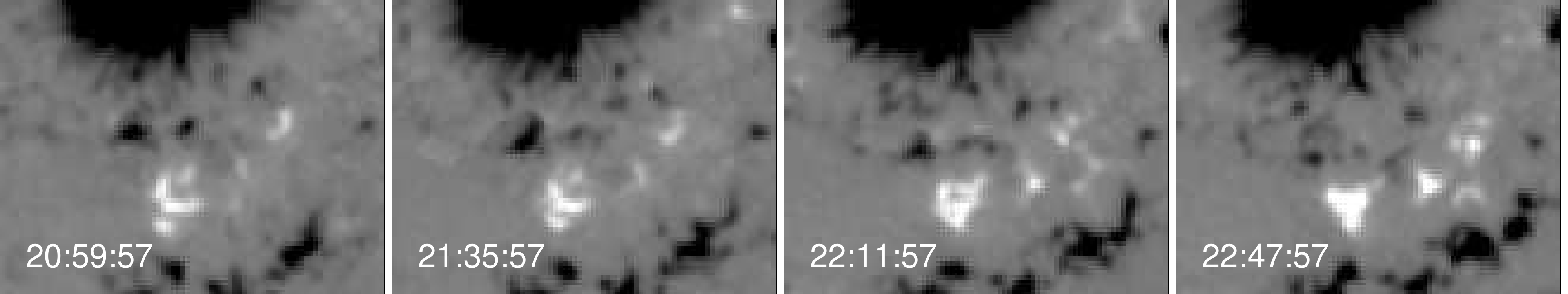}
  \caption{Time sequence of HMI magnetograms shows a small emerging
    positive polarity under the west lobe of the BPSS.}
  \label{fig:fluxemergence}
\end{figure*}

\section{Study of the Magnetic Field}
\label{sec:topol}

In our previous work \citep{Jiang2012c, Jiang2016F}, we have developed a
data-constrained CESE--MHD model to reconstruct the 3D coronal magnetic
field. The MHD model, based on the space-time conservation-element and
solution-element (CESE) scheme, is designed to focus on the
magnetic-field evolution and to consider a simplified solar atmosphere
in gravity with a small plasma $\beta$. Magnetic vector-field data
derived from the observations at the photosphere is inputted directly
to constrain the model. With regarding that for the studied event the
coronal configuration changes are very limited and can be approximated
by successive MHD equilibria, we have solved a time sequence of MHD
equilibria basing on a set of vector magnetograms taken by HMI around
the time of flare. The model and results have been given in details in
\citep{Jiang2012c}, which shows that the model successfully reproduces
the basic structures of the 3D magnetic field, as supported by the
good visual similarity between the field lines and the coronal loops
observed by AIA. Here our analysis is confined to the magnetic
topology and its change related with local region of the flare site.

% how to compute the BP
%With a squashing-fator map, we find that there is a BP co-spatial in
%the flare location.

Even before inspecting the coronal field, we find there is indeed BPs
on the photosphere.  With the vector magnetogram, one can identify BPs
directly by searching for locations satisfying
\begin{equation}
  \label{eq:BP}
  B_{z} = 0 \ \
  {\rm and}\ \
  \vec B\cdot\grad B_{z} > 0,\ \
  z = 0 \ \ ({\rm i.e.,\ on\ the\ photosphere}).
\end{equation}
This equation characterizes portions of PILs where the horizontal
field crosses from the negative to positive $B_{z}$. {\it SDO}/HMI has
taken a sequence of vector magnetograms with cadence of 12~min. Here
we select the magnetogram taken at 22:12~UT, see
\Fig~\ref{fig:AIA_BP}~(a), which is mostly near the time of the
flare. Locating of the BPs is carried out on a mesh refined by ten
times from the original magnetogram using linear interpolation, and
then we search for all the grid points fulfilling
Equation~(\ref{eq:BP}).  As shown in \Fig~\ref{fig:AIA_BP}~(c), the BP
points are plotted as black diamond shapes overlaid on the AIA-1600
image (the grid coordinates are co-aligned with the the AIA
image). Here it can be seen that the BP is clearly co-spatial with the
flare location, mainly near flare patch A. The BPs mainly reside on
the long PIL in the central region of the flare site (the thick cyan
line on \Fig~\ref{fig:AIA_BP} (b)), and there are some other small
segments of BPs near this main PIL. On the main PIL, the BPs likely
belong to a continuous long BP line, although they are disconnected
frequently as derived from the observed vector field due to the noise
of the magnetogram. One may worry about the reliability of the data
since in this region the vector fields is relatively weak ($\sim 200$~G), but
considering that the directions of the magnetic vectors do not show
the pattern of random noise, we can be confident with the existence of
these BPs.

We note that no BP coincides with the flare patches B and C. Are
this places related with the footpoints of the BPSS field lines?
%To answer this, we need to study the coronal magnetic field reconstructed by the MHD model.
%The AIA observations appears to approve this conjecture and to
%confirm it, we explore
With the 3D coronal field model, we trace all the field lines passing
through the BP points, which forms the BPSS, and locate the footpoints
of all these BPSS field lines. \Fig~\ref{fig:AIA_BP} (b) shows all
these BPSS field lines, and their footpoints are marked as blue
diamonds, which are also overlaid on the AIA-1600 image. As we have
conjectured, the western BPSS footpoints are co-spatial very well with
the flare ribbon B and C, while the eastern footpoints coincides with
flare patch A.  A 3D view of the BPSS field lines is also shown in
\Fig~\ref{fig:BP3D}, from which we can see that the apexes of the
field lines are $3\sim 4$~Mm, thus the whole structure is very
low. The field lines outline the topology of the separatrix surface
which separates three different topological regions. The footpoints in
the east are very near the BPs, thus their brightenings along with
those from the BPs produce the big flare patch A that consists of many
even-small flare patches. Moreover, there should be another separatrix
surface (not shown here) at the west of the BPs, which divides the
west part into two connectivity domains and results in two different
sets of footpoints, thus produces the two flare patches B and C,
respectively. Furthermore, comparing \Fig~\ref{fig:AIA_BP} (b) with
(d) shows that the shape of the BPSS is in good agreement with the hot
flaring loop observed in AIA-131, strongly suggesting that
reconnection occurs in this BPSS and heats the plasma in corresponding
field lines to form the hot loops.  Besides,
%there appears to be similarity of shapes between the field lines with
the mini-filament as aforementioned (shown by arrows in \Fig~\ref{fig:flarelocal}) might be supported at the dips of field lines just above the BP
separatrix surface. The disappearance
of this mini-filament after the flare seems to suggest that the
dipped field lines are lift up by magnetic reconnection at the BPSS,
which make the field lines detached from the photosphere at the BP,
and consequently the dense plasma is drained off by gravity.
  However, there is also a possibility that the disappearance of the
  mini-filament is because it was heated to a higher temperature
  insensitive to AIA-171 and 304 passbands.

% Our preliminary study shows that there is a BP co-spatial with this
% flare site \citep{Jiang2012apj}.  These observations suggest that
% these brightening kernels are associated with the BP and its
% separatrix footpoints, as we will confirm in the following.

\begin{figure*}[htbp]
  \centering
  \includegraphics[width=\textwidth]{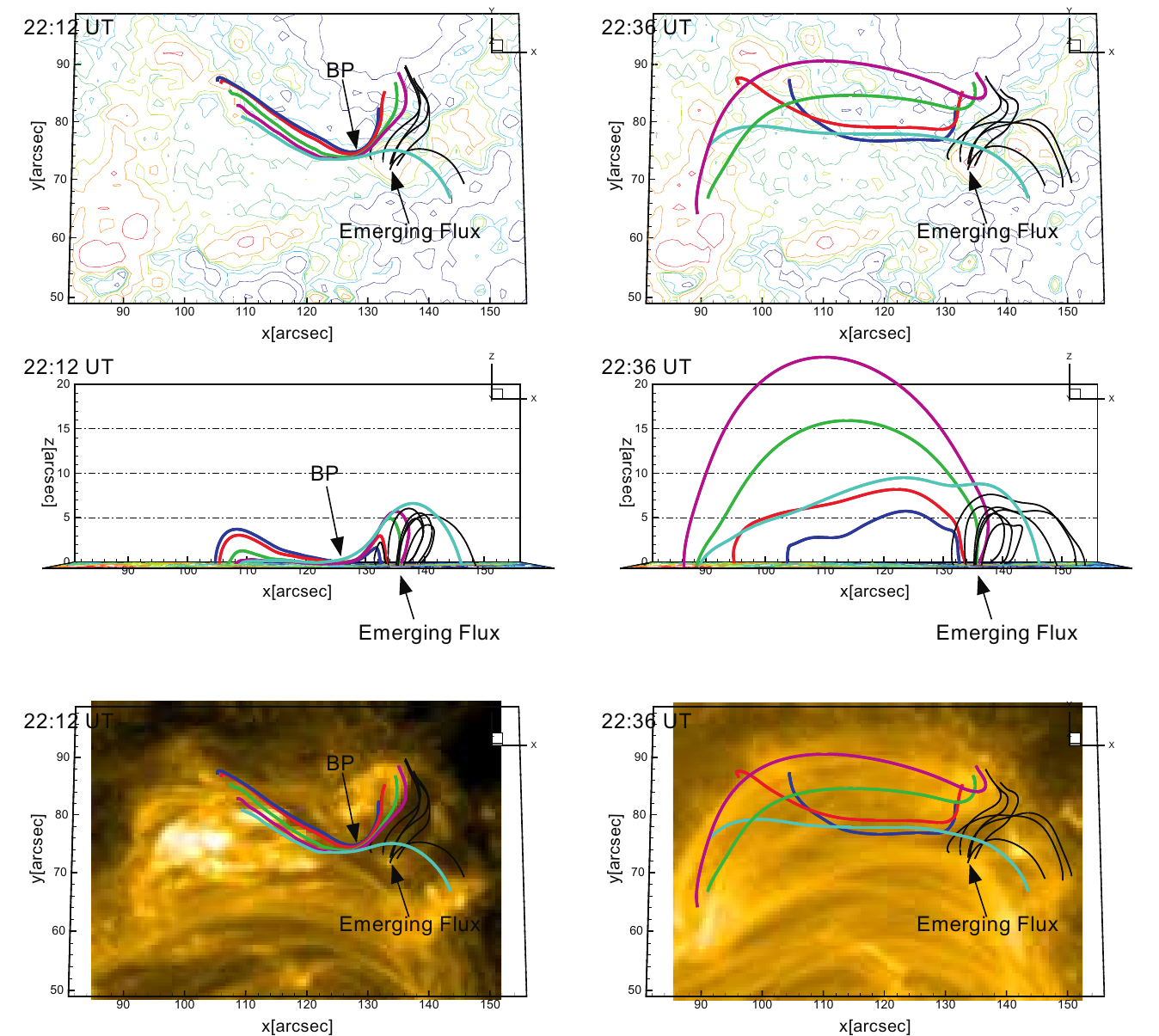}
  \caption{Magnetic configuration change from pre-flare state (left panels)
    to post-flare state (right panels). The magnetic field lines are shown in
    SDO view angle (top) and side view (middle), and overlaid on
    AIA-171 images (bottom). A group of BPSS field lines are shown in thick
    colored lines. Their footpoints in the west (the right side) are
    fixed for the pre-flare and post-flare field lines. A group of
    field lines from the emerging polarity are shown in black
    color. The contour lines represent photospheric magnetogram
    $B_z$, red for positive and blue for negative.}
  \label{fig:BPliftup}
\end{figure*}

With the above study based on the field at the time of 22:12~UT, we
have given a strong evidence that the flare is co-related with the BP,
and the process of the event is suggested as following. Firstly
magnetic reconnection is activated at the BP, then the accelerated
particles there traced the field lines near the BP-separatrix and
produced flare ribbons, and the field lines that attached the
photosphere at the BP before the flare are lift up by the reconnection
and expand upward, manifesting by the emergence of the post-flare
loops and the disappearance of the mini-filament. Such and similar
processes can usually be related with flux emergence, which can
trigger reconnection in the BPSS, for example in the formation of
active region \citep{Pariat2004} and in the surges and arch filament
systems \citep{Mandrini2002}. In the present event, by inspecting the
photospheric field evolution, we find that there is indeed an emergence
of a small positive polarity under the west lobe of the BPSS, see
\Fig~\ref{fig:fluxemergence}. With the evolution of the reconstructed
magnetic field we are able to illustrate how the BPSS field lines
evolves through the flare. The left panels of \Fig~\ref{fig:BPliftup}
show several of BPSS field lines (the colored thick lines) before the
flare, while the right panels show their post-flare
configuration. Under the west lobe of the BPSS, the emerging positive
polarity is shown with a set of field lines (the black thick
lines). The field lines rooted in the emerging flux expand upward (as
comparing the post-flare state with the pre-flare state) and push the
overlaying BPSS, which continuously increases electric currents in the
BPSS and free energy to the system. When the current sheet is enhanced
enough for resistivity to be important, magnetic reconnection sets in
at the BPSS, and the field lines attached at the BP by the photosphere
plasma are lift up, expand as they relax, forming the post-flare
loops. The field lines might also undergo slipping reconnection as
they expand, with the footpoints not necessarily being in the same
pre-flare locations. In the bottom panels, the field lines are
overplotted on the AIA-171 images, which demonstrates a nice agreement
of the model field lines with the coronal loops.

\section{Conclusions}
\label{sec:conc}

In this study, we analyzed a C2.3 confined flare in AR~11117 with SDO
observation and data-constrained MHD models. The flare is a non-standard
one consisting of multiple ribbons, and it occurs in a region
between the AR sunspots where the magnetic field is relatively weak
and no distinct magnetic shear is observed, although the AR contains a
typical fast-emerging and strong-field site with significantly sheared
PIL. With strong evidences, we concluded that this flare is triggered
by magnetic reconnection in a BPSS and can be identified as a BP
flare:
\begin{enumerate}

\item By direct inspecting the
photospheric magnetic vector field measured by SDO/HMI, we find there
are BPs on the PILs matching parts of the flare
ribbons.

\item From the 3D coronal magnetic field derived from the MHD
model constrained by the vector magnetograms, we find strikingly good
agreement of the BPSS footpoints with the flare ribbons, and the BPSS
itself with the hot flaring loop system.

\item Moreover, the triggering of
the BP flare can be attributed to a small flux emergence under the
lobe of the BPSS, and the relevant change of the coronal magnetic field is
well reproduced by the pre and post-flare MHD solutions which
matches the corresponding pre and post-flare AIA observations. The
flux emergence under the lobe of the BPSS triggers reconnection at the
BP, making the pre-flare field lines that attached the photosphere at the BP
relax and expand upward and form the post-flare
loops.
\end{enumerate}

Our work contributes to the study of non-typical flares that constitute the majority of solar flares but cannot be explained by the
standard flare model.  It is also worth noting that here the flare scenario
is shown by MHD results
constrained by observation data,
% rather than cartoon-like conjectures which are traditionally
%used in analysis of flare observations.
and such data-constrained or even data-driven MHD
modeling the evolution of coronal magnetic field is becoming an
new important approach for gaining a comprehensive understanding of
the physical nature of non-typical flares~\citep{Jiang2016NC}.

\normalem
\begin{acknowledgements}

  This work is jointly supported by National Natural Science
  Foundation of China (41531073, 41374176, 41574170, 41231068, and
  41574171) and the Specialized Research Fund for State Key
  Laboratories. Data from observations are courtesy of NASA {SDO}/AIA
  and the HMI science teams

\end{acknowledgements}

%\bibliographystyle{raa}
%\bibliography{all}

\end{document}